# Selection of the optimal embedding positions of digital audio watermarking in wavelet domain

Yangxia Hu[1,*], Maode Ma[1,2], Wenhuan Lu[1], Neal N. Xiong[3], Jianguo Wei[1,4]

[1] College of Intelligence and Computing, Tianjin University, Tianjin 300350, China
[2] School of Electrical and Electronic Engineering, Nanyang University of Technology, Singapore
[3] Department of Mathematics and Computer Science, Northeastern State University, Tahlequah, OK 74464, USA
[4] School of Computer Science and Technology, Qinghai Nationalities University, Xining 810007, China

* Corresponding author: huyangxia@tju.edu.cn

*Abstract*—This work studied embedding positions of digital audio watermarking in wavelet domain, to make beginners understand the nature of watermarking in a short time. Based on the theory of wavelet transform, this paper analyzed statistical distributions of each level after transformation and the features of watermark embedded in different transform levels. Through comparison and analysis, we found that watermark was suitable for embedding into the coefficients of the first four levels of wavelet transform. In current state-of-art approaches, the embedding algorithms were always to replace the coefficient values of the embedded positions. In contrast this paper proposed an embedding algorithm of self-adaptive interpolation to achieve a better imperceptibility. In order to reduce the computational complexity, we took a pseudo random sequence with a length of 31 bits as the watermark. In the experiments, watermark was embedded in different locations, including different transform levels, high-frequency coefficients and low-frequency coefficients, high-energy regions and low-frequency regions. Results showed that the imperceptibility was better than traditional embedding algorithms. The bit error rates of the extracted watermark were calculated and we analyzed the robustness and fragility of each embedded signal. At last we concluded the best embedding positions of watermark for different applications and our future work.

## I. Introduction

### A. Motivations

ELECTRONIC watermarking can be traced back as far as 1954[1]. In the next few decades, the wider applications of multimedia products in the open internet have made more and more digital products illegally copied and maliciously tampered. To solve these problems, more and more researchers began to study electronic watermarking and digital watermarking did not receive substantial interest as a research topic until the 1990's [2,3]. At first, digital watermarking was mainly used in the world of graphics and later was gradually used in audio signal because human auditory perception system was more sensitive than visual perception system [4]. In [5], the author introduced some watermark algorithms used in images and pointed out the application of digital audio watermarking. In the next thirty years, digital watermarking technology has been rapidly developed and more and more widely used, and the mechanism of watermarking has been more and more optimized. For different carriers, different watermarks and different applications, corresponding algorithms have been proposed and widely applied [6,7].

The main process of watermarking includes watermark embedding and extracting. In the process of embedding, the embedding position directly affects the complexity of algorithm, the characteristics of watermark, and application scenarios of watermarked signal. Therefore, choosing the best embedding positions for different watermarks is an important step [8,9]. Research on embedding positions can help beginners understand the fundamental principles of watermarking in a short time, which is a foundation for the further research of watermark generation, embedding and detection. A good embedding position can not only balance the imperceptibility, robustness and embedding capacity of watermark, but also reduce the complexity of the algorithm [10,11]. However, there is no literature that systematically introduces and analyzes the optimal embedding positions of watermark, and no researcher studies and discusses the embedding positions of watermark in detail. In literatures, authors usually point out the embedding positions of watermark directly and focuses on the embedding algorithms, which makes beginners feel hard to learn.

Currently there are countless articles on different applications of watermarking, such as copyright disputes and authentication of integrity, robust watermarks and fragile watermarks, watermark embedding, detection and extraction, spatial domain watermarking, transform domain watermarking, and hybrid domain watermarking. Watermarking technology has developed for decades, and is becoming more and more perfect [6,8,12,13].

The selection of watermark embedding positions involves a lot of knowledge of signal processing. Spatial domain watermarking is simple in calculation and has a larger embedded capacity, but its security is relatively poor, which is easy to be cracked by criminals [14-17]. Transform domain watermarking is gradually widely used because of its better imperceptibility and security. Discrete Wavelet Transform (DWT) is widely used in the field of images at first, and gradually applied to audio signals in recent years [18-20]. Embedding the watermark in wavelet domain can not only make the watermark evenly distributed to the frequency domain coefficients, but also analyze the time domain and frequency domain characteristics of the signal in real time [21-23].

When we study the best embedding positions of watermark in wavelet domain according to different applications, we must select out wavelet bases, the number of wavelet transform levels, low-frequency or high-frequency coefficients, and regions with high-energy or low energy [24-26]. These problems are the premise and foundation of watermark embedding, and it is usually very difficult for beginners, who do not understand the essence of watermarking. Only by mastering the knowledge of watermarking process can we adopt different embedding mechanisms and get watermarked audio signals with different applications.

### B. Main Contributions

Our main contributions are as follows.
1) We introduced the principle of wavelet transform, process of DWT in digital watermarking, and discussed the effect of watermark with different wavelet bases. Then we illustrated selection of different decomposition levels, embedding regions with different frequencies and energies through theories and experiments.
2) We proposed an embedding algorithm based on adaptive interpolation, which did not replace the coefficient values of the embedding position with watermark, but interpolated watermark values into the corresponding position to change the statistical distributions of the coefficients before and after embedding as little as possible.

The paper continued as follows. In Section II, we discussed related work. Section III gave the related theories and technology. In Section IV, we introduced an embedding algorithm of self-adaptive interpolation. In section V, we gave our experimental results and analysis. Conclusions and future work were described in section VII.

## II. RELATED WORK

None of previous work focused on selection of watermark embedding positions, and previous researches were always about effects of watermark after embedding. Such as different embedding domains, different generation and embedding mechanisms, different recovery and detection algorithms, improvement of robustness, and so on.

Technology of digital watermarking has developed fast during recent thirty years. At first, watermark was embedded in spatial domain of original signal, for example, Least Significant Bit (LSB). Lu *et al.* [27] introduced a kind of fragile watermark which could recover tampered speech, and watermark was embedded in the LSB of original signal. Javier *et al.* [28] proposed a fragile watermarking scheme for color-image authentication and self-recovery, in which original image was divided into non-overlapping blocks, and watermark generated by each block was embedded within the 2-LSB. Gradually, watermark was embedded in transform domains and mixed domains, such as DWT, Discrete Cosine Transform (DCT), and DWT-DCT, to achieve a better imperceptibility or robustness. Wu [29] introduced audio digital watermarking schemes in transform domain in detail, which included DWT, DCT, and DWT-DCT. Mosleh *et al.* [16] proposed a scheme that the original audio signal was first split into set of segments, and then each segment was taken DCT transform. After then, DCT coefficients were framed and watermark was embedded in frequency sub-band. Sunesh *et al.* [30] proposed a blind digital watermarking method with DCT in order to achieve the security, imperceptibility, robustness, blindness, unambiguity, and simplicity. Niu *et al.* [31] proposed a blind audio watermark decoder in Stationary Wavelet Transform (SWT) domain based on bivariate generalized Gaussian Distributions, and they experimentally confirmed that the proposed approach performed well compared to the state-of-the-art audio watermarking methods. In recent ten years, time domain watermark algorithms were proposed and widely used. Li [32] introduced an algorithm that watermark was embedded in time domain by Spread Spectrum technology in his doctoral dissertation, and he verified that the algorithm had a larger embedding capacity and a smaller computational complexity. The disadvantage of this method was that it was only suitable for copyright protection, and compared with embedding watermark in transform domains, the latter was more widely used. Another watermarking method was based on compressed domains, since compressed versions of audio files were popular on the internet. Qiao *et al.* [33] proposed embedding watermark into audio streams which were in MPEG format directly. Scheme of embedding was simple but it was not secure enough and embedded watermark was easily detected out.

The idea of wavelet transform was first proposed by French engineer J. Morlet in 1974. In 1986, the famous mathematician Y. Meyer constructed the first wavelet base, and then it began to be widely used [34]. Compared with Fourier Transform, wavelet transform was a local transformation of time and frequency, so it could extract information from signals effectively. It inherited and developed the idea of localization of short-time Fourier transform, and provided a time-frequency window which overcame the disadvantage that the size of window did not change with frequency. It was an ideal tool for time-frequency analysis and processing of signal, and had been widely used in the study of digital watermarking [35,36]. In recent years, more and more researchers are processing signals in DWT domains to study algorithms of watermark embedding, extraction and detection. Hu *et al.* [17] proposed a blind watermarking method in DWT domain. Lee *et al.* [24] presented a blind speech watermarking algorithm which embedded watermark information into separate DWT sub bands. Diego *et al.* [37] presented a new fragile watermarking method for digital audio authenticity to give a high degree of fragility, and the selected component to embed the secret key was the approximation of DWT. Li *et al.* [26] proposed a scheme using the norm ratio of approximate coefficients in DWT. Original signal was framed and DWT was applied to each frame. Toshiki *et al.* [38] concluded that different wavelet filters provided different watermarking performances based on the wavelet transform, and proposed a new scheme based on DWT. Hu *et al.* [35] proposed a scheme jointly exploring the Rational Dither Modulation (RDM) and auditory masking properties in the DWT domain to achieve effective blind audio watermarking. Huang *et al.* [21] proposed a new blind digital audio watermarking system by using optimization-based modification on low-frequency amplitude of DWT. Mehdi *et al.* [39] proposed a novel high capacity robust audio watermarking algorithm by using the high frequency band of DWT. Attari *et al.* [23] proposed a blind and robust audio watermarking scheme by using Fibonacci numbers properties as well as DWT, which embedded watermark bits in low-frequency coefficients of DWT to make watermark be less sensitive. Lei *et al.* [40] introduced a robust watermarking algorithm based on DWT, and he combined DWT with DCT and Singular Value Decomposition (SVD). Watermarked signal had a good imperceptibility and strong robustness to common audio signal processing attacks. Liu [41] introduced a blind digital audio watermarking method based on vector quantization in his doctoral dissertation, in which watermark was embedded in DWT domain and watermarked signal had a better imperceptibility and robustness. Tan *et al.* [42] proposed a robust channel coding-based watermarking scheme which embedded watermark in 4-level DWT and could achieve near exact watermark recovery against all kinds of attacks.

Selection of appropriate wavelet bases is always a problem for beginners. The commonly used wavelet bases are Haar wavelet, Daubechies (db$N$) wavelet, Morlet wavelet, Meyer wavelet [43]. Different wavelet bases have different effects on transform levels, since their different vanishing moments, support length, and so on.

In different applications, watermark is usually embedded in different transform levels, since the higher the decomposition level is, the more different between the characteristics of coefficients obtained and the original signal are. By analyzing the statistical characteristics and spectrum characteristics of high-frequency and low-frequency coefficients, we can know the changes of the signal before and after transformation of each level.

Researchers have analyzed coefficients of transform domains many years ago. Chen *et al.* [44] analyzed parameters of modified DCT domain. For DWT, we know that a low-frequency coefficient and a high-frequency coefficient are generated after each level of wavelet transform. The low-frequency coefficients contain most of the energy of the signal before transformation and include the overall characteristics of the audio signal. The energy of the high-frequency coefficients is small, which include detailed features of the audio signal [26]. Watermarked signal has different performances when watermark is embedded in low-frequency coefficient and high-frequency coefficient. The common parameters which can describe the characteristics of audio signal are energy and zero crossing rate. The energy distributions of high-frequency and low-frequency coefficients after wavelet transform are uneven, and the effect is different when different energy regions are selected [45]. Watermarked signal has a better robustness when embedding watermark in the region with higher energy, and it has a better fragility when embedding watermark in the region with lower energy.

Watermark embedding and extraction are the most important process in digital watermarking system. Watermark generation is the premise of watermark embedding. Researchers often choose different watermarks according to different requirements. Wang et al. [14] introduced a watermark scheme which took a binary image after Arnold transformation as watermark to verify the robustness of watermarked signal. Lu et al. [27] took the compressed version of original signal as watermark to recover the tampered signal. These watermarks are meaningful, but in some applications, we do not need to know the content of watermark and achieve our goals by comparing the watermark before and after embedding. These watermarks are meaningless. Cui et al. [25] analyzed the application of wavelet coefficients in digital watermarking and took a chaos sequence as watermark information. Marzieh et al. [46] introduced a watermark detection scheme in which they took a random sequence as watermark.

In the recent thirty years, many watermark embedding algorithms have been proposed according to different applications. When the performance of the watermarking system gradually met the needs of the market, researchers began to propose simpler and more targeted algorithms. Adaptive embedding algorithms were proposed, and embedding positions and algorithms could be adjusted according to the characteristics of coefficients. Originally, these methods were called watermarking technology base on audio content. Huang [47] proposed an audio watermarking method based on audio content in his doctoral dissertation, which pointed out that embedding watermark according to the energy, zero crossing rate, and some other characteristics of coefficients. Dutta et al. [48] proposed an efficient watermarking algorithm which embedded watermark data adaptively in the audio signal. The degree of embedding was adaptive in nature and was chosen in a justified manner according to the localized content of the signal. Lei et al. [15] proposed an optimal and secure audio watermarking scheme based on self-adaptive particle swarm optimization and quaternion wavelet transform. Chen et al. [20] proposed an adaptive audio watermarking via the optimization point of view on the wavelet-based entropy. Hu [49] introduced an Adaptive Vector Norm Modulation (AVNM) scheme to achieve a satisfactory balance of imperceptibility, robustness, and payload capacity. Gupta et al. [36] proposed an efficient and fast watermarking scheme based on Lifting Wavelet Transform (LWT) and quantization. The sum in a group of the approximation coefficients is quantized using a quantization parameter, determined adaptively. Youssef [50] proposed a novel Hybrid Fuzzy Self-Adaptive Digital Audio Watermarking scheme (HFSA-AW) in which embedded watermark in the original signal was self-adaptive. Hu et al. [24] introduced an adaptive mean modulation scheme which increased the embedding capacity of watermark. In recent years, adaptive methods are being used in many signals processing systems and becoming more and more popular. Wu et al. [51] proposed an improved version of diffusion Gauss-Newton, which was adaptive to sudden changes on noisy range measurements and could locate target in industrial environments.

Existing adaptive algorithms always changed the values at embedding positions. In fact, we expect the change is minimal before and after embedding, and watermarked signal has a better imperceptibility. In this paper, we proposed a watermark embedding algorithm of self-adaptive interpolation, which made the process of watermark embedding and extraction simpler, watermarked signal had a better imperceptibility, and was more suitable for studying the optimal embedding positions.

We can know the distribution differences of audio coefficients before and after DWT of different levels by analyzing their waveforms and spectrogram. Other characteristics, such as energy and zero crossing rate, can be obtained by calculation. Kaur et al. [52] introduced how to calculate the energy and analyze zero crossing rate. The distribution characteristics of audio signal can be observed by statistical model, such as Probability Density Function (PDF) and Cumulative Distribution Function (CDF). Chakrabarti et al. [53] analyzed the statistical characteristics of the coefficients of audio signal after wavelet transform. In the state of art studies, statistical characteristics were often used in watermark detection. Sun et al. [54] proposed that detection of digital watermarking could be a question of a binary hypothesis test, and generalized Gaussian distributions were applied to statistically model the DCT coefficients of the original image, and the detector had a good performance. Marzieh et al. [46] proposed a scheme based on analyzing statistical characteristics of coefficients. Etemad et al. [55] designed a new multiplicative watermark detector by statistical modeling. Ivana et al. [56] presented a watermark embedding procedure based on PDF modeling, which ensured optimal detection for weak watermarks even when they were exposed to some common attacks such as noise addition and compression. B. Gunsel et al. [57] introduced a statistical framework for audio watermark detection and decoding, which modeled the statistics of watermarked and original audio signals by Gaussian Mixture Models (GMM) with $K$ components. Niu et al. [58] proposed a novel digital image watermark decoder in the Nonsubsampled Shearlet Transform (NSST) domain, wherein a PDF based on the bivariate was used. At the watermark receiver, the statistical model parameters of bivariate Weibull distribution were estimated effectively.

Although researchers studied digital watermarking in different transform domains, it was known that watermarking technology was most commonly used in wavelet domain. In order to make our study more widely used, we chose to study the best embedding positions in wavelet domain. The problems to be considered in the embedding process included the selection of wavelet basis functions, the number of transformation layers, high-frequency or low-frequency coefficients, high or low energy. We would introduce the selection criteria in different application scenarios through theory and experiment in this paper. In addition, in order to simplify the embedding algorithm and make the watermark imperceptible, we proposed an adaptive interpolation embedding scheme and took pseudo-random sequence as watermark information. Through reading the literatures, it was easy to know that statistical analysis was also widely used in the watermark system. In our experiments, we analyzed the statistical characteristics of the signal to judge the changes of the signal. We hope that beginners could have a deeper understanding of the nature of watermarking through this paper, and they could do Subsequent deeper researches.

III. RELATED THEORIES AND TECHNOLOGY

## A. The Principle of Wavelet Transform

Wavelet transform evolved from Fourier transform. At first, Fourier transform was used to decompose and reconstruct signal, and it made the signal composed of several things which could make it better than before after processing and restoration. The amount of decomposition, which was the base, was just like a vector. The base of Fourier transform was a sine curve of different frequencies, so Fourier transform decomposed a signal wave into superposition sum of sine waves with different frequencies. The calculation process of Fourier transform was shown in Equation (1).

$$F(w) = \int_{-\infty}^{\infty} f(t) * e^{-iwt} dt, \quad (1)$$

In (1) $w$ was frequency. The base of Fourier transform covered the whole-time domain, as (a) in Fig.1 showed, and we could see that Fourier transform could not describe the local characteristics of the signal in time domain, and it was not suitable for analyzing abrupt and non-stationary signals. Most of audio signals in our life were non-stationary, therefore, the idea of wavelet transform was proposed and widely used in processing of non-stationary signals. The idea of wavelet transform directly changed the base of Fourier transform into a wavelet base of finite length, as (b) in Fig.1 showed, which was attenuated. In this way, not only the frequency domain information could be obtained, but also the time domain information could be located. The calculation process of wavelet transform was as Equation (2) showed.

$$WT(a, \tau) = \frac{1}{\sqrt{a}} \int_{-\infty}^{\infty} f(t) * \Psi\left(\frac{t-\tau}{a}\right) d_t. \quad (2)$$

It could be seen from (1) and (2) that Fourier transform has only the variable of frequency ω, however, wavelet transform had two variables, scale $a$ and translation amount $\tau$. Scale $a$ controlled the expansion and contraction of the function, and the translation amount $\tau$ controlled the translation of the function. The scale corresponded to the frequency, and the translation amount $\tau$ corresponded to the time. In this way, we could know not only the frequency component of the signal, but also its position in time domain. Therefore, wavelet transform provided a theoretical foundation for processing non-stationary signals.

## B. Some Commonly Used Wavelet Bases

Haar wavelet was one of the earliest compactly supported orthogonal wavelet bases, and its wavelet function was relatively simple. It was a single rectangular wave in the supported domain of $t \in [0,1]$. Haar wavelet was defined as Equation (3).

$$\Psi(t) = \begin{cases} 1 & 0 \le t \le 1/2, \\ -1 & 1/2 < t \le 1, \\ 0 & else. \end{cases} \quad (3)$$

It could be seen from (3) that Haar wavelet was easy to calculate and constituted a group of orthogonal normalized wavelets. Waveforms of time domain and frequency domain were shown in Fig. 2.

Daubechies wavelet was constructed by Ingrid Daubechies, who was a famous wavelet analysis scholar in the world. It was usually abbreviated as db*N*. *N* was the vanishing moment of wavelet, and the supported length was *2N-1*. It was often used as a filter. Waveforms of db4 wavelet in time domain and frequency domain and its four filters were shown in Fig.3. We could see from (a) that db4 wavelet was not symmetrical. In the process of decomposition and reconstruction, filters were symmetrical and in opposite directions. Morlet wavelet had no scale function and its decomposition was non orthogonal and Meyer wavelet was not compactly supported and converged quickly. Haar wavelet and db*N* wavelet were commonly used in digital watermarking systems.

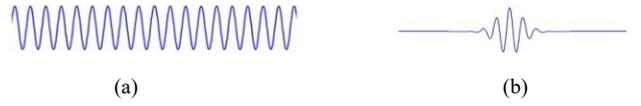

(a)  (b)

Fig.1. Changes of bases. (a) Base of Fourier transform. (b) Base of wavelet transform.

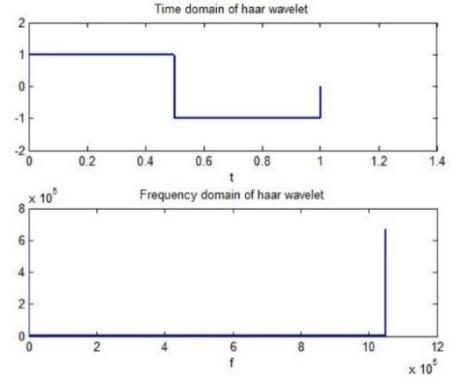

Fig.2. Waveforms of time domain and frequency domain of Haar wavelet. It was symmetrical, orthogonal and discontinuous in time domain. When the frequency reached a certain value, it would change suddenly.

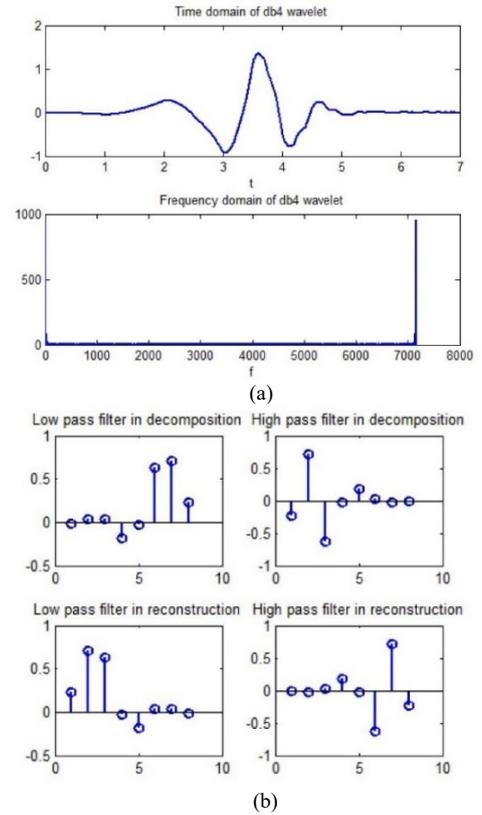

Fig. 3. Characteristics of db4 wavelet. (a)Waveforms in time domain and frequency domain. (b) Four filters of db4 wavelet.

## C. Process of DWT

The 3-level discrete wavelet transform process of one-dimensional signal was represented in Fig.4 [26].

As shown in Fig.4, $h_0^{(n)}(n)$ and $h_1^{(n)}(n)$ represented the low-pass filter and high-pass filter respectively in each level of transform. Then, a low-frequency coefficient and a high-frequency coefficient were obtained after down sampling with a scale of 2. The time domain and frequency domain characteristics of high-frequency and low-frequency coefficients could be observed in real time. By analyzing the statistical characteristics and spectrum characteristics of high-frequency and low-frequency coefficients, we could know the changes of the signal before and after transformation of each level.

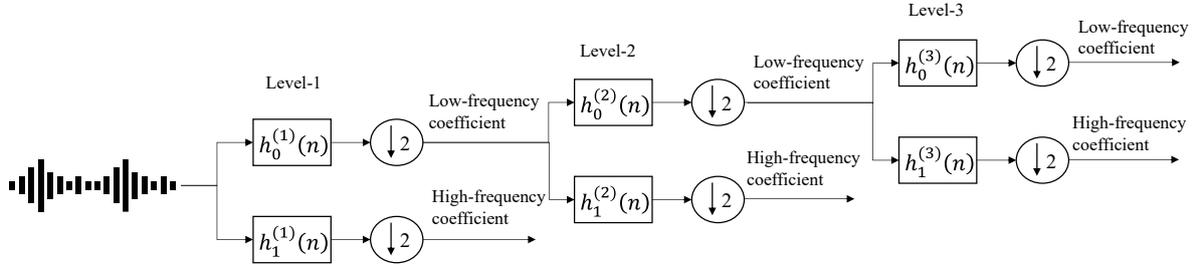

Fig.4. The 3-level DWT

## IV. An Embedding Algorithm of Self-adaptive Interpolation

With the development of watermarking technology, more and more self-adaptive embedding algorithms were proposed, which improved the performance of watermark compared with the regular algorithms [59]. In current state-of-art approaches, the embedding algorithms were always to replace the coefficient values with the watermark to embed, however, in our experiment, we proposed an embedding algorithm of self-adaptive interpolation. It was different from the traditional embedding algorithm that this algorithm did not replace the coefficient values of the embedding position with watermark, but interpolated watermark values into the corresponding position to change the statistical distributions of the coefficients before and after embedding as little as possible. At the same time, the watermark information we embedded was a pseudo-random sequence, which reduced the calculation process of watermark preprocessing [59], and was suitable for studying the optimal embedding position for embedding the watermark.

### A. Watermark Embedding

The overall embedding scheme of watermark in our experiments was shown in Fig.5.

Fig. 5 showed the general process of watermark embedding, where $L$ was the length of each audio segment after segmentation, $w$ was the watermark sequence and $k$ was the embedding key. In the experiment, $k$ was the primitive polynomial of the pseudo random sequence generator, and $\alpha$ was the embedding strength. In order to simplify the process of watermark preprocessing and obtain the optimal embedding positions, the value of $\alpha$ in the experiment was 1. The detailed watermark embedding process could be described as Algorithm 1 showed.

Algorithm 1 presented the embedding process after 1-level DWT. Watermark was embedded in the region with high energy of low-frequency coefficient.

In experiments, the length of pseudo-random sequences was determined by the length of original signal and the number of segmentations. We could divide our proposal into following three parts.
1) Preprocessing: Remove the silent segments of input signal to reduce calculation load. Then divide new signal into $n$ segments.
2) Embedding: Generate pseudo-random sequences accor-

**Algorithm 1:** Embedding
1 Input signal $S_0$;
2 $sil$=find($S_0(i)$==0);
3 remove $S_0(i)$==0 → $S_0'$;
4 segmentation $S_0'$ → $S_1, S_2,\cdots, S_n$;
5 generate pseudo-random sequences of length $n$→ $P$;
6 **foreach** $S_i$ **do**
7     1-level DWT → $L_1$ and $H_1$;
8     $index$=find(max(abs($L_1(j)$)));
9     **if** $index$==1
10     $L_1$'= $P(i)$+ $L_1$;
11     $index_0$=find(min(abs($L_1(j)$)));
12     remove $L_1$'($index_0$) →$S_{Li}$';
13     **else**
14     $L_1(index$-1)= $P(i)$;
15     others= $L_1$→$L_1$';
16     remove $L_1$'($index_0$) →$S_{Li}$';
17     1-level IDWT to $S_{Li}$' and $H_1$→ watermarked $S_i$';
18     **end**
19 connect $S_1$', $S_2$',$\cdots$, $S_n$' → watermarked $S$';
20 connect $S$' with $sil$→ watermarked signal $Sw$.

ding to the length of tested signal. Then do $n$-level DWT to each segment, and $n$ was determined by the application. Choose low-frequency or high-frequency and calculate the maximal and minimal energy and embed watermark according to obtained index.
3) Postprocessing: Do IDWT to each embedded low-frequency coefficients and high-frequency coefficients, and then connect each segment and add silent segments. Then watermarked signal was obtained.

In order to describe the embedding process more clearly, we gave the flow chart of embedding process, as Fig.6 showed.

Fig.6 showed the process of watermark embedding after 1-level transform. The colorful rectangular blocks in Fig.6 represented the corresponding coefficients and watermark values. The imperceptibility of watermark was obtained by Equation (4) [22].

$$I(s(t),w(t)) = 10 * log_{10}(\sum_t s^2(t) / \sum_t w^2(t)), \quad (4)$$

In (4), $I$ represented imperceptibility, $s(t), w(t)$ represented the original signal and the watermarked signal respectively. It was easy to know that the closer $I$ to 0 was, the closer the energy of two signals were, and the better the imperceptibility was.

### B. Watermark Extraction

Watermark extraction was the inverse process of watermark embedding. Since our proposal was blind, we did not need original signal while extracting. The detailed process was as Algorithm 2 showed.

As Algorithm 2 showed, in our experiment, pseudo-random sequences were composed of 0 and 1, and we removed silent s-egments while the silent sequence was successive to avoid deleting watermark information. When we extracted the watermark sequence $P(1), P(2),\cdots, P(n)$, we could compare it with generated sequence by the same primitive polynomial $k$ in embedding process, and calculate the bit error rate.

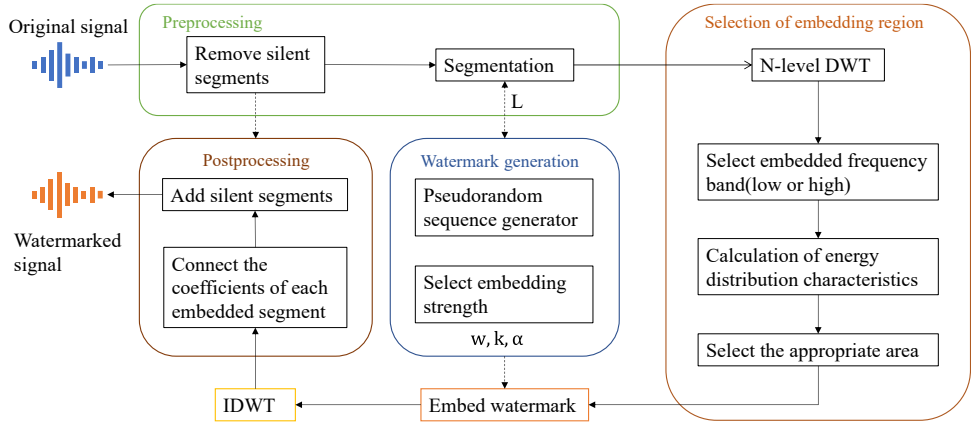

Fig. 5. The embedding process of watermark

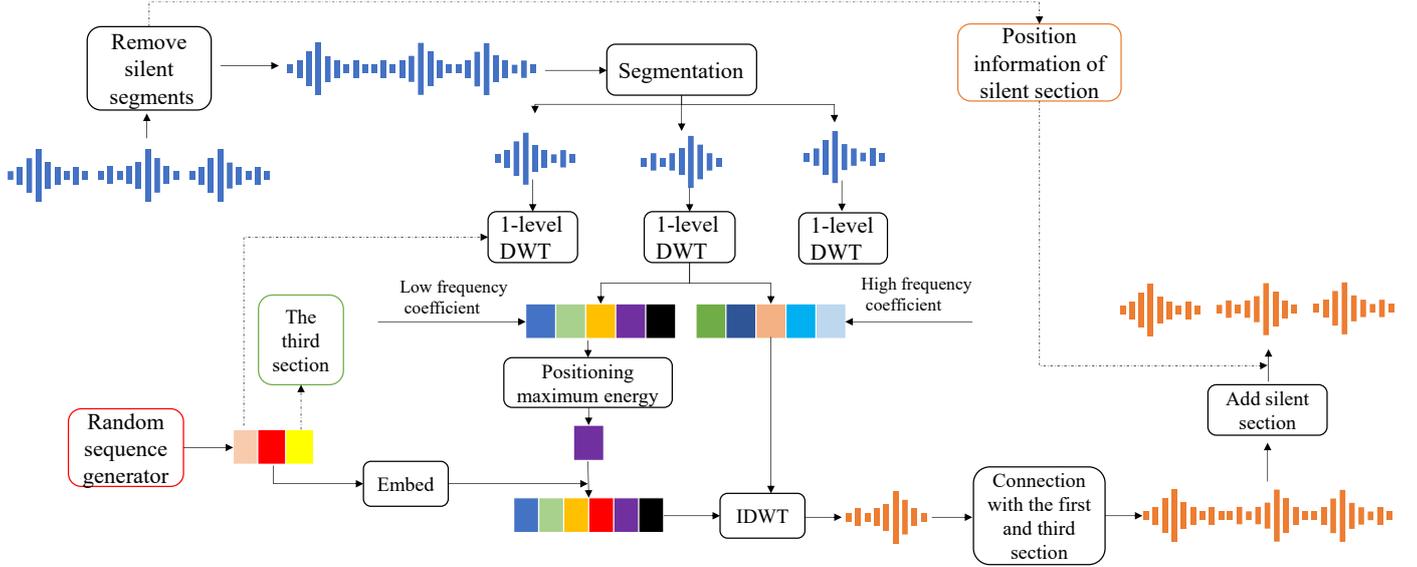

Fig. 6. The detailed watermark embedding process of 1-level DWT. This figure represented processes of segmentation, 1-level DWT, selection of embedding position and embedding. The dark purple rectangle represented coefficients with the maximum energy, and the red rectangle represented watermark value to embed.

**Algorithm 2:** Extraction
1 Input watermarked signal $Sw$;
2 $sil$=find($Sw(i)$==0);
3 remove successive $Sw(i)$==0 → $Sw'$;
4 segmentation $Sw'$ → $Sw_1, Sw_2,\cdots, Sw_n$;
5 **foreach** $Sw_i$ **do**
7    1-level DWT → $L_1$ and $H_1$;
8    $index$=find(max(abs($L_1(j)$)));
9    **if** $index$==1
10     $error$;
11    **else**
12     $P(i)$=$L_1$($index$-1);
13    **end**

## V. EXPERIMENTAL RESULTS AND ANALYSIS

### A. Selection of Wavelet Bases

In order to analyze and compare the differences between wavelet bases on watermark systems, six different styles of audio signal were selected in the experiment, which were Speech (in Chinese), Blues, Classic, Rock, Jazz, Pop. Each audio signal was in wav format, and the length of each audio was 20s, and the sampling frequency was 16000 Hz. Haar wavelet and db$N$ wavelet were selected as wavelet bases, and $N$ was from one to four. The distribution characteristics and changes of high-frequency and low-frequency coefficients of each level after wavelet transformation were analyzed by using statistical toolbox of MATLAB. It was found that the distributions of high-frequency and low-frequency coefficients of each level obtained by different wavelet bases were almost the same. The analysis results of Speech after 1-level transformation of by using different wavelet bases were shown in Table I.

TABLE I
CHARACTERISTICS OF HIGH-FREQUENCY AND LOW-FREQUENCY COEFFICIENTS OF SPEECH AFTER 1-LEVEL TRANSFORMATION BY USING DIFFERENT WAVELET BASES

| Wavelet bases | Mean value of high frequency | Mean value of low frequency | Variance of high frequency | Variance of low frequency |
|---|---|---|---|---|
| Haar | 0.00000279889 | 0.00108466 | 0.00201349 | 0.197818 |
| db1 | 0.00000279889 | 0.00108466 | 0.00201349 | 0.197818 |
| db2 | 0.00000279889 | 0.00108463 | 0.0020725 | 0.199441 |
| db3 | 0.00000279889 | 0.00108466 | 0.00200739 | 0.199568 |
| db4 | 0.00000279889 | 0.00108456 | 0.00205319 | 0.199597 |

From Table I, we could see the means and variances of high-frequency coefficients and low-frequency coefficients after 1-level wavelet transform had almost no difference between different wavelet bases. We knew that in digital watermarking systems, the following work would not involve further operations of signal transformation. As a result, the selection of these 5 wavelet bases could be arbitrary and it was not the main factor affecting the watermark effect. In the following experiments, we chose Haar wavelet as the base to study the optimal embedding positions.

*B. Selection of Wavelet Transform Levels*

In the experiment, the wavelet toolbox of MATLAB was used to do ten levels of wavelet transform for these six audio segments. The sampling frequency of each audio signal was 16000 Hz. The spectrogram of high-frequency and low-frequency coefficients of each level after transformation were observed respectively, and the distribution characteristics of high-frequency and low-frequency coefficients of each level were analyzed with the statistical toolbox of MATLAB. The statistical results of the high-frequency and low-frequency coefficients of the first ten levels of Speech and Blues were shown in Table II.

The solid color areas of the distribution characteristics in Table II were the histograms of the original signal and the high-frequency and low-frequency coefficients after decomposition, and we could see that with the increase of decomposition level, the color became lighter and lighter, which indicated that the density of coefficients was getting smaller and smaller. The colored lines were their distributions of probability densities. The red line corresponded the original signal, the blue line corresponded the low-frequency coefficients, and the brown line corres-

TABLE II
(a) DISTRIBUTIONS OF COEFFICIENTS IN EACH LEVEL OF SPEECH

| Decomposition level | Total energy (o)174.91 | Distribution characteristics |
|---|---|---|
| 1 | 193.94 | 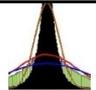 |
| 2 | 205.89 | 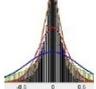 |
| 3 | 219.52 | 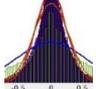 |
| 4 | 221 | 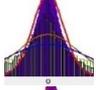 |
| 5 | 182.26 | 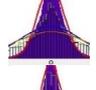 |
| 6 | 100.61 | 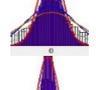 |
| 7 | 39.684 | 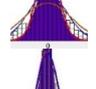 |
| 8 | 22.699 | 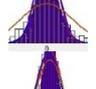 |
| 9 | 10.438 | 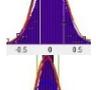 |
| 10 | 5.9984 | 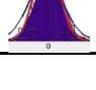 |

TABLE II
(b) DISTRIBUTIONS OF COEFFICIENTS IN EACH LEVEL OF BLUES

| Decomposition level | Total energy (o)136.9 | Distribution characteristics |
|---|---|---|
| 1 | 136.79 | 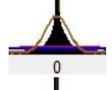 |
| 2 | 131.19 | 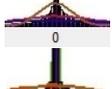 |
| 3 | 123.54 | 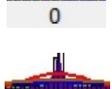 |
| 4 | 110.28 | 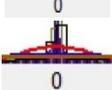 |
| 5 | 87.602 | 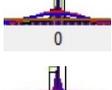 |
| 6 | 55.935 | 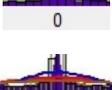 |
| 7 | 36.023 | 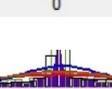 |
| 8 | 21.877 | 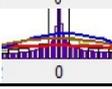 |
| 9 | 9.0997 | 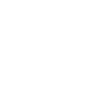 |
| 10 | 7.2801 | 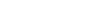 |

ponded the high-frequency coefficients. The "(o)" below "Total energy" was the total energy of original signal [15].

By analyzing their probability density distribution curves, we could see that the distributions of the first four levels of transform were consistent for Speech and Blues. In level-1, the distribution of low-frequency coefficient was very similar to the original signal, and they were almost the same. In level-2, The distribution of low-frequency coefficient was closer to original signal. In level-3, the distribution of high-frequency coefficient was closer to original signal, and they were almost the same. In level-4, the distribution of high-frequency was closer to original signal. After more than five levels, the distributions of frequency coefficients after decomposition were different between different audio segments. By calculating the energy of each level after transformation, it was found that the energy of the first four levels was closer to the original energy. When decomposition level was higher than five, the energy was sharply reduced. Spectrogram of the low-frequency coefficients of each level of Speech were shown in Fig. 7.

The spectrogram showed that the first four layers could clearly show the characteristics of the original signal. The higher level of transformation was, the more blurred the spectrogram was, and it appeared disordered lines. The results showed that the high-frequency and low-frequency coefficients obtained by the first four levels of transformation could represent the characteristics of the original signal and were suitable for embedding watermark.

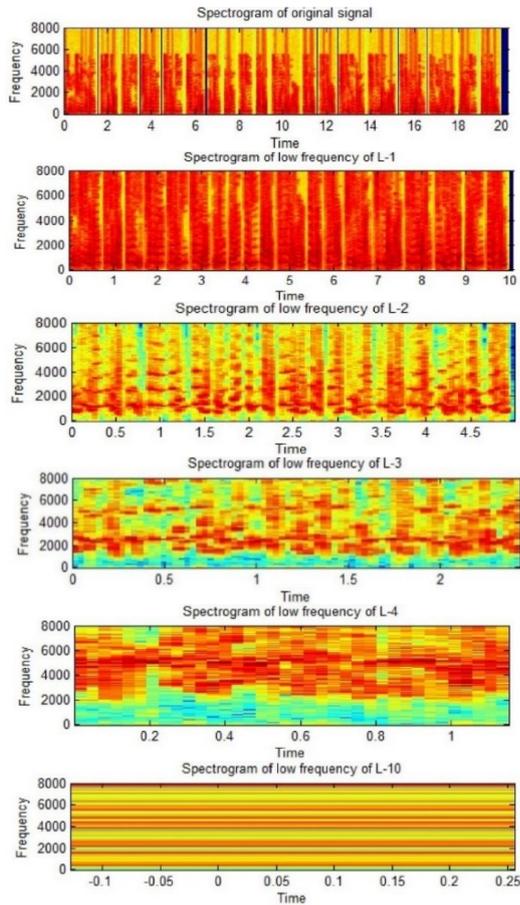

Fig. 7. Spectrogram of the low-frequency coefficients of each level of Speech

### C. Selection of High Frequency and Low Frequency

From Fig.4, we knew that a low-frequency coefficient and a high-frequency coefficient were generated after each level of wavelet transform. The low-frequency coefficients contained most of the energy of the signal before transformation and included the overall characteristics of the audio signal. The energy of the high-frequency coefficients was small, which included detailed features of the audio signal. The result of Speech after 1-level wavelet transform with the base of Haar was shown in Fig.8.

In Fig.8, $a_1$ and $d_1$ represented the waveforms of low-frequency coefficient and high-frequency coefficient respectively after wavelet transform. Through comparison, it could be seen that the amplitude of low-frequency coefficient was close to the original signal, and they were almost the same, which indicated that the low-frequency signal contained most of the energy of the original signal, while the amplitude of the high-frequency coefficient was much smaller, which indicated that the energy of the high-frequency signal was very small and contained the detail components of signal.

In the experiment, watermark was embedded in the place where the energy of low-frequency coefficient and high-frequency coefficient were higher. The watermark information was still a pseudo-random sequence with the length of 31 bits, and the sampling frequency was 16000 Hz. The embedding algorithm of adaptive interpolation was used and we analyzed and calculated the imperceptibility by Equation (4). The experimental results of imperceptibility were shown in Table III.

From Table III, we could see that the imperceptibility of embedding high-frequency and low-frequency coefficients in each level was different. The imperceptibility of embedding low-frequency coefficients in the second and forth levels was better than that in the first and third levels, while the imperceptibility of embedding high-frequency coefficients in the first and third levels was better than that in the second and fourth levels. Then we analyzed and calculated robustness by bit error rates after extracting. Results of Speech were shown in Table IV.

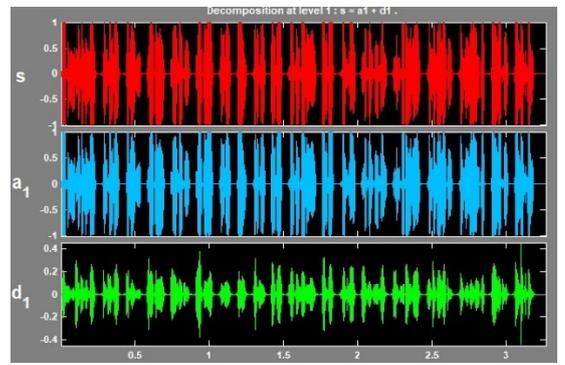

Fig. 8. Result of Speech after 1-level wavelet transform with the base of Haar

Table IV showed the results of extraction of the watermark embedded in high-frequency and low-frequency coefficients with larger energy after Lossy compression (MP3-16k), in which "×" indicated that watermark extracted was wrong, and "-" indicated that the index value was 1 in the extraction process, so the watermark at the corresponding position could not be extracted. After embedding the high-frequency coefficients after transform of level 2-4, the extracted results were all not correct. In the same way, we did experiments of Noise addition, Random cropping, Re-sampling, Re-quantization, Lowpass filtering, the low-frequency results were shown in Table VIII. However, the watermark extracted in high-frequency coefficients were almost all not correct. The results showed that watermark embedded in high-frequency coefficients could not resist external attacks and could be embedded as fragile watermark.

### D. Selection of High Energy and Low Energy

The common parameters which could describe the characteristics of audio signal were energy and zero crossing rate. The energy distributions of high-frequency and low-frequency coefficients after wavelet transform were uneven, and the effect was different when different energy regions were selected. In the watermarking system, we usually chose the corresponding embedding region by calculating the energy of the signal. In the experiment, watermark was embedded in the high-energy region and low-energy regions of 6 audio segments after 4 levels of wavelet transform. The sampling frequency of each audio was 16000 Hz. The imperceptibility of the embedded watermark signal was calculated by Equation (4), and the robustness was observed and analyzed. The experimental results of imperceptibility were shown in Table V.

It could be seen from Table V that watermark embedded in the high-energy regions of the second and fourth levels was more imperceptible, and watermark embedded in low-energy regions of the first and third levels was more imperceptible. After attack experiment, the robustness of the watermark embedded in high-energy regions was shown in Table VIII, while the watermark embedded in the low-energy regions could not resist the attack, and the extracted watermark sequence was not correct. Therefore, the watermark embedded in low-energy regions was suitable to be used as fragile watermark to verify integrity.

### E. Imperceptibility Test

In the experiment, Haar wavelet was used to embed watermark into six audio segments, and the sampling frequency of each audio was 16000 Hz. Watermark was embedded in the region with higher energy of low-frequency coefficients. The results

TABLE III
VALUES OF *I* OF LOW-FREQUENCY AND HIGH-FREQUENCY COEFFICIENTS OF 6 TYPES OF AUDIO

| Number of embedded levels | Types of audio | | | | | | | | | | | |
|---|---|---|---|---|---|---|---|---|---|---|---|---|
| | Speech | | Blues | | Classic | | Rock | | Jazz | | Pop | |
| | Low frequency | High frequency | Low frequency | High frequency | Low frequency | High frequency | Low frequency | High frequency | Low frequency | High frequency | Low frequency | High frequency |
| 1 | 0.065022 | 0.0038412 | -0.021824 | 0.0038412 | -0.020324 | 0.0038412 | 0.021824 | 0.0038412 | 0.065824 | 0.0038412 | -0.020824 | 0.0038412 |
| 2 | 0 | 0.024699 | 0 | 0.024699 | 0 | 0.024699 | 0 | 0.024699 | 0 | 0.024699 | 0 | 0.024699 |
| 3 | 0.091399 | 0.080556 | 0.091399 | 0.080556 | 0.091399 | 0.080556 | 0.091399 | 0.080556 | 0.091399 | 0.080556 | 0.091399 | 0.080556 |
| 4 | 0.14059 | 0.15226 | 0.14059 | 0.15226 | 0.14059 | 0.15226 | 0.14059 | 0.15226 | 0.14059 | 0.15226 | 0.14059 | 0.15226 |

TABLE IV
THE RESULTS OF SPEECH

| | Index | 1 | 2 | 3 | 4 | 5 | 6 | 7 | 8 | 9 | 10 | 11 | 12 | 13 | 14 | 15 | 16 | 17 | 18 | 19 | 20 | 21 | 22 | 23 | 24 | 25 | 26 | 27 | 28 | 29 | 30 | 31 |
|---|---|---|---|---|---|---|---|---|---|---|---|---|---|---|---|---|---|---|---|---|---|---|---|---|---|---|---|---|---|---|---|---|---|
| | Original watermark sequence | 1 | 0 | 0 | 0 | 0 | 1 | 1 | 0 | 1 | 0 | 1 | 0 | 0 | 1 | 0 | 0 | 0 | 1 | 0 | 1 | 1 | 1 | 1 | 1 | 0 | 1 | 1 | 0 | 0 | 1 | 1 |
| Extracted values | Lossy compression (MP3-16k) (Level 1) | Low frequency | | | | | | | | | | | | | | | | | | | | | | | | | | | | | | |
| | | 1 | 0 | 0 | 0 | 0 | × | 1 | 0 | 1 | 0 | 1 | 0 | 0 | 1 | 0 | 0 | 0 | 1 | 0 | 1 | 1 | 1 | 1 | 1 | 0 | 1 | 1 | 0 | 0 | 1 | 1 |
| | | High frequency | | | | | | | | | | | | | | | | | | | | | | | | | | | | | | |
| | | 1 | - | - | - | - | - | - | - | - | - | 1 | - | 0 | 1 | 0 | - | 0 | 1 | 0 | 1 | 1 | 1 | - | 1 | | 1 | - | 0 | 0 | - | - |

TABLE V
VALUES OF *I* OF DIFFERENT ENERGY REGIONS OF 6 TYPES OF AUDIO

| Number of embedded layers | Types of audio | | | | | | | | | | | |
|---|---|---|---|---|---|---|---|---|---|---|---|---|
| | Speech | | Blues | | Classic | | Rock | | Jazz | | Pop | |
| | High energy | Low energy | High energy | Low energy | High energy | Low energy | High energy | Low energy | High energy | Low energy | High energy | Low energy |
| 1 | 0.065022 | 0.0019208 | -0.021824 | 0.0019208 | -0.020324 | 0.0030113 | 0.021824 | 0.0019412 | 0.064289 | 0.0038412 | -0.020824 | 0.0030415 |
| 2 | 0 | 0.024356 | 0 | 0.024356 | 0 | 0.024699 | 0 | 0.024599 | 0 | 0.024699 | 0 | 0.024699 |
| 3 | 0.091399 | 0.089091 | 0.091399 | 0.089091 | 0.091399 | 0.089506 | 0.091399 | 0.080506 | 0.091399 | 0.080556 | 0.091399 | 0.080456 |
| 4 | 0.14059 | 0.14475 | 0.14059 | 0.14475 | 0.14059 | 0.15226 | 0.14059 | 0.14226 | 0.14059 | 0.15226 | 0.14059 | 0.14236 |

of imperceptibility were shown in Table VI.

It could be seen from Table VI that all the values of *I* obtained by the second level were 0, which indicated that watermark embedded in the second level of wavelet transform had the best imperceptibility. Wang Xiang-Yang used an embedding algorithm which replaced coefficients with watermark in [59], and this was the most common method. The results of imperceptibility calculated by [59] were shown in Table VII.

Table VII showed the imperceptibility of the watermark obtained by the embedding algorithm mentioned in [59], which replaced the coefficient with watermark. In order to compare it with the interpolation embedding algorithm in this paper, we used the same audio signal in the experiment, and embedded the same watermark in the same position. We could see that absolute values of *I* in Table VI were bigger than those in Table VII, which meant that our algorithm had a better imperceptibility than the traditional embedding algorithm.

### F. Experiments of The Capability to Resist Attacking

Experiments including Noise addition, Random cropping, Re-sampling, Re-quantization, Lowpass filtering, and Lossy compression, were carried out on watermarked signals of each level. Then watermark information was extracted and bit error rate was obtained. The specific experimental process was as follows and the results of Speech were shown in Table VIII.

1) Noise addition: Gaussian white noise with mean value of 0 and variance of 0.01 was added to the watermarked signal.
2) Random cropping: Shear 10% of watermarked signal randomly.
3) Re-sampling: Set the sampling frequency of watermarked signal from 16kHz to 8kHz and then to 16kHz, and from 16kHz to 44.1kHz and then to 16kHz.
4) Re-quantization: Watermarked signal was quantized from 16 bits to 8 bits and then re-quantized to 16 bits.
5) Lowpass filtering: Lowpass filters with cut-off frequency of 2kHz, 3.2kHz, 6kHz were adopted.
6) Lossy compression: Watermarked signal was compressed into MP3 format of 16k.

From Table VIII, we could see that for attacks of Noise addition, Random cropping, Re-sampling (16k-8k-16k), and Lowpass filtering (6kHz), values of BERs were increasing with the increase of transform levels, and more than 0.5 at 3-level transform. For attacks of Re-sampling (16k-44.1k-16k) and Re-quantization (16bit-8bit-16bit), values of BERs were more than 0.5 at 4-level transform. For attacks of Lowpass filtering (2kHz and 3.2kHz), values of BERs were all above 0.5, which meant that watermarked signal could not resist these two kinds of attacks. For attacks of lossy compression (MP3-16k), values of BERs were all under 0.3, which meant that watermarked signal had a better robustness to lossy compression.

Therefore, in order to make watermark resist attacks of cropping and noise, we should select the first two levels. For the re-sampling and re-quantization attacks, we should also select the first two levels to receive a better robustness. For lowpass filtering attack, watermark had no robustness when cut-off frequency was far away from sampling frequency. And our watermark could resist attack of lossy compression.

TABLE VI
VALUES OF *I* OBTAINED BY OUR PROPOSAL

| Number of embedded level | Types of audio | | | | | |
|---|---|---|---|---|---|---|
| | Speech | Blues | Classic | Rock | Jazz | Pop |
| 1 | 0.065022 | -0.021824 | -0.020324 | 0.021824 | 0.065824 | -0.020824 |
| 2 | 0 | 0 | 0 | 0 | 0 | 0 |
| 3 | 0.091399 | 0.091399 | 0.091399 | 0.091399 | 0.091399 | 0.091399 |
| 4 | 0.14059 | 0.14059 | 0.14059 | 0.14059 | 0.14059 | 0.14059 |

TABLE VII
VALUES OF *I* OBTAINED BY [59]

| Number of embedded level | Types of audio | | | | | |
|---|---|---|---|---|---|---|
| | Speech | Blues | Classic | Rock | Jazz | Pop |
| 1 | 0.069423 | -0.024523 | -0.021512 | 0.023105 | 0.067811 | -0.023136 |
| 2 | 0.000124 | 0.000078 | 0.000124 | 0.000124 | 0.000124 | 0.000124 |
| 3 | 0.092571 | 0.092571 | 0.092571 | 0.092571 | 0.092571 | 0.092571 |
| 4 | 0.142042 | 0. 142042 | 0. 142042 | 0. 142042 | 0. 142042 | 0. 142042 |

TABLE VIII
THE RESULTS OF SPEECH

| Types of attacks | BERs of different levels | | | |
|---|---|---|---|---|
| | 1 | 2 | 3 | 4 |
| Without attack | 0 | 0 | 0 | 0 |
| Noise addition (0,0.01) | 0.09 | 0.16 | 0.55 | 0.9 |
| Random cropping by 10% | 0.16 | 0.26 | 0.58 | 0.84 |
| Re-sampling (16k/8k/16k) | 0.06 | 0.35 | 0.52 | 0.68 |
| Re-sampling (16k/44.1k/16k) | 0.06 | 0.23 | 0.39 | 0.65 |
| Re-quantization (16bit/8bit/16bit) | 0.03 | 0.19 | 0.35 | 0.58 |
| Lowpass filtering (2kHz) | 0.55 | 0.9 | 1 | 1 |
| Lowpass filtering (3.2Hz) | 0.84 | 0.84 | 1 | 1 |
| Lowpass filtering (6kHz) | 0.29 | 0.39 | 0.68 | 0.84 |
| Lossy compression (MP3-16k) | 0.03 | 0.09 | 0.26 | 0.29 |

*G. Industrial Applications*

Experimental results illustrated effect of watermarked signal when different embedding positions were selected. Readers could have a deeper understanding of the nature of watermarking in a short time. The proposed embedding algorithm had a better imperceptibility than common proposals. The results of this paper can be applied in the following aspects.
1) Balance the relationship among imperceptibility, watermark capacity and robustness in transform domain, since researchers could select optimal position and adjust parameter to make watermarked signal have better performance.
2) By embedding different positions, robust watermarking for copyright protection and fragile watermarking for integrity authentication can be obtained.
3) Proposed embedding algorithm had a simpler calculation, which was suitable for large sample capacity and watermark information was meaningless sequences, For example, watermark detection.
4) Selection of optimal embedding position is the premise of watermark embedding, which is suitable for all watermarking systems.

VI. CONCLUSIONS AND FUTURE WORK

From the analysis and experiments, we got following rules for the selection of wavelet bases, embedding level numbers, high-frequency or low-frequency coefficients, and high- energy or low-energy regions, which could be a reference for watermarking learners in the first stages.
1) Different wavelet bases had little effect on experimental results.
2) Watermark was suitable for embedding in the first four levels of wavelet transform.
3) Watermark embedded in low-frequency coefficients of level-2 transformation had the best imperceptibility.
4) Low-frequency coefficients were suitable for embedding robust watermarks, while high-frequency coefficients were suitable for embedding fragile watermarks.
5) The imperceptibility and robustness of watermarks were better when regions with higher energy were selected.
6) It was suitable for embedding fragile watermarks in the regions with lower energy.
7) The robustness of watermark decreased with the increase of transform levels.
8) The designed algorithm had better imperceptibility than previous embedding algorithms and could resist lossy compression attacks and other common attacks when watermarks were embedded in lower levels transform.

Beginners can refer to the above suggestions to choose the embedding position of watermark. In addition, researchers can improve the watermark system from the following aspects, which are also our future work.

1) It is necessary to design more complicated embedding mechanisms to study more complicated attacks [60].
2) Mixed transform domains can improve the security of watermark and can be applied to future research, such as DWT-DCT and dual tree complex wavelet transform.
3) Study watermark detection of audio signal with large sample size.
4) Explore new applications of watermarking systems.

## ACKNOWLEDGEMENTS

This research was supported by the National Natural Science Foundation of China (No. NSFC61876131), and the Key Basic Research and Development of Ministry of Science and Technology (No. 2018YFC0806802).